\documentclass[prl,aps,reprint,amsmath,amssymb,longbibliography, superscriptaddress]{revtex4-2} 

\usepackage[dvipsnames]{xcolor}
\usepackage[colorlinks=true,urlcolor=blue,citecolor=purple,linkcolor=red]{hyperref}
\usepackage{amssymb,amsmath,amsfonts}
\usepackage{epsfig}
\usepackage{graphicx}
\usepackage{epstopdf}
\usepackage{natbib}
\usepackage{braket}

\usepackage{lmodern}
\usepackage[T1]{fontenc}

\def\clock{{\count0=\time
		\divide\count0 60
		\ifnum\count0<10 0\fi\the\count0
		\multiply\count0 -60 \advance\count0 \time
		:\ifnum\count0<10 0\fi \the\count0
}}
\newcommand{\timestamp}{{\small\vbox{\hbox{\tt\jobname.tex}
			\hbox{\the\day/\the\month/\the\year, \clock}}}}

\hypersetup{%
  pdftitle   = {Chiral anomaly from (anomalous) spin hydrodynamics},
  pdfkeywords = {Anomalous hydrodynamics, holography with R-currents, spin currrents},
  pdfauthor  = {Jay Armas, Giorgos Batzios},
}

\newcommand{\C}{\mathbb{C}}

\newcommand{\be}{\begin{eqnarray}}
\newcommand{\ee}{\end{eqnarray}}
\newcommand{\beq}{\begin{eqnarray}}
\newcommand{\eeq}{\end{eqnarray}}

\newcommand{\beqa}{\begin{eqnarray}}
\newcommand{\eeqa}{\end{eqnarray}}

\usepackage{mathrsfs}

\renewcommand{\u}{\boldsymbol{u}}
\ifdefined\C
\renewcommand{\C}{\boldsymbol{C}}
\else
\newcommand{\C}{\boldsymbol{C}}
\fi

\definecolor{gris}{rgb}{0.5,0.5,0.5}
\definecolor{darkgreen}{rgb}{0.0,0.5,0.0}

\allowdisplaybreaks[0]
\begin{document}

	\title{Chiral anomaly from (anomalous) spin hydrodynamics}

	\author{Jay Armas}
	\email{j.armas@uva.nl}
	\affiliation{Institute for Theoretical Physics, University of Amsterdam, 1090 GL Amsterdam, The Netherlands}
    \affiliation{Dutch Institute for Emergent Phenomena, The Netherlands}
 \affiliation{Institute for Advanced Study, University of Amsterdam, Oude Turfmarkt 147, 1012 GC Amsterdam, The Netherlands}
 \affiliation{Niels Bohr International Academy, The Niels Bohr Institute, University of Copenhagen, Blegdamsvej 17, DK-2100 Copenhagen \O{}, Denmark}

	\author{Giorgos Batzios}
	\email{g.batzios@uva.nl}
	\affiliation{Institute for Theoretical Physics, University of Amsterdam, 1090 GL Amsterdam, The Netherlands} 
    \affiliation{Dutch Institute for Emergent Phenomena, The Netherlands}

	\begin{abstract}
We show that the low energy fluctuations of spinning black Dp branes are described by a theory of spin hydrodynamics on a spacetime  $\mathbb{M}_{p+1}\times \mathbb{T}^{n+2}$ in which the fluid is flowing on $\mathbb{M}_{p+1}$ and spinning on $\mathbb{T}^{n+2}$. Focusing on the hydrodynamic regime of $\mathcal{N}=4$ supersymmetric Yang-Mills theory, we provide a geometric interpretation of the R-current anomaly in terms of a gravitational anomaly from the ten-dimensional point of view. This follows from the holographic duality between a spinning fluid in ten dimensions and an anomalous chiral fluid in four dimensions. We comment on the relations between the theory of spin hydrodynamics introduced here and other theories of spin hydrodynamics in the context of heavy-ion collisions. \\ \center{\emph{Dedicated to the memory of Umut G\"{u}rsoy}}.
          \end{abstract}
	
\maketitle	

 The chiral anomaly and spin hydrodynamics \cite{Huang:2015oca, Kharzeev:2024zzm, Becattini:2024uha, Florkowski:2024cif} constitute two, seemingly different, areas of intense research in both high-energy and condensed matter physics. In the context of electroweak interactions, the chiral anomaly is responsible for the rapid pion decay rate \cite{PhysRev.177.2426, Anomaly}, while in heavy-ion collisions it is expected to give rise to novel transport properties of the quark-gluon plasma such as the chiral vortical and chiral magnetic effects \cite{Huang:2015oca, Kharzeev:2024zzm}. At the same time, the observation of spin polarization of $\Lambda$ hyperons in heavy-ion collisions suggests that spin degrees of freedom should be included in hydrodynamic descriptions \cite{Becattini:2024uha}. In condensed matter physics, the chiral anomaly leads to unusual electronic properties such as negative magnetoresistance \cite{weyl} whereas spin currents can be measured in liquid metals \cite{liquidmetal}. In this letter, we describe a connection between chiral transport and spin hydrodynamics in the context of holography. More precisely, we uncover a duality between hydrodynamics with \emph{transverse} spin in ten spacetime dimensions and an anomalous chiral fluid in four spacetime dimensions \footnote{In fact this duality is present between anomalous chiral fluids in four spacetime dimensions and spinning fluids in six spacetime dimensions. However, since we want to highlight the string theory origin of these chiral fluids we consider spinning fluids in ten spacetime dimensions.}.
 
Large classes of 4d strongly coupled plasmas when sourced by background electromagnetic fields are expected to exhibit a chiral anomaly \cite{Landsteiner:2016led}. Indeed, upon gauging, $\mathcal{N}=4$ supersymmetric Yang-Mills theory (SYM) has an R-current anomaly \cite{Witten:1998qj}. In the low-energy, hydrodynamic regime of $\mathcal{N}=4$ SYM, holography has revealed the existence of chiral transport determined by the presence of this $U(1)^3$ anomaly \cite{Erdmenger:2008rm, Banerjee:2008th, Son:2009tf}. However all of these studies come from a 5d perspective, making use of dimensional reduction on the $S^{5}$, and consider fluctuations of 5d R-charged black holes \cite{Cvetic_1999, Erdmenger:2008rm, Banerjee:2008th, Son:2009tf}. Here we uplift the rich physics of chiral transport to 10d in which the low energy effective theory of spinning Dp-branes plays a crucial role.

We show that this low energy description is a theory of spin hydrodynamics on a spacetime $\mathbb{M}_{p+1}\times \mathbb{T}^{n+2}$ in which the fluid lives on the worldvolume $\mathbb{M}_{p+1}$ and spins on the transverse space $\mathbb{T}^{n+2}$. For the case of the near-horizon limit of the D3 brane, the transverse angular momenta (or spins) corresponding to the 3 transverse Cartan planes of rotation realize holographically the charges of the $SO(6)$ R-symmetry group of the dual field theory. After introducing the local current whose integral provides the global spin charges, which we call the spin current, we explain how the hydrodynamics of the R-charged black holes of \cite{Erdmenger:2008rm, Banerjee:2008th, Son:2009tf} corresponds to the hydrodynamics of (anomalous) spinning fluids characterising the long-wavelength fluctuations of spinning black D3 branes. This enables a geometrization of the chiral anomaly of the 4d theory in terms of a gravitational anomaly in 10d. Nevertheless, this is merely an artifact of imposing a specific dimensional reduction ansatz, and in the full 10d string theory picture this anomaly is absent. We argue that the precise geometric origin of the R-current anomaly should be understood via the coupling of the worldvolume fluid with the five-form flux, leading to the mapping of the chiral fluid in 4d to a (non-anomalous) spinning fluid with higher-form charge in 10d. We begin by looking at the thermodynamics and currents of spinning D3 branes in order to extract the low energy description of their dynamics.

\vspace{0.2em}
\noindent
\textit{Currents of spinning D3 branes from gravity---} We consider asymptotically flat black brane solutions of type IIB supergravity with worldvolume $\mathbb{M}_{p+1}$ spinning on the transverse $\mathbb{T}^{n+2}$ space in $d=n+p+3$ spacetime dimensions. It is always possible to introduce a coordinate $\u=1/r$ such that when $\u\to0$ and hence $r\to\infty$ a far away observer with respect to the black brane horizon sees a distribution of matter that admits a multipole expansion of the form \cite{Armas:2011uf}
\begin{equation} \label{eq:stress}
 \bold T^{\mu\nu}=T^{\mu\nu}(\sigma)\hat \delta(X)-\nabla_\rho\left(T^{\mu\nu\rho}(\sigma)\hat \delta (X) \right)+...~~,  
\end{equation}
where $T^{\mu\nu}(\sigma)$ is the monopole stress-tensor, $\sigma^{a}$ coordinates on $\mathbb{M}_{p+1}$, $T^{\mu\nu\rho}$ the dipole moment of the stress-tensor, $\delta (X)=(\sqrt{-\gamma}/\sqrt{-g})\delta^{(n+2)}(x^\alpha-X^\alpha)$ the weighted $(n+2)$ dimensional delta function where $x^\mu$ are coordinates on spacetime with metric $g_{\mu\nu}$, determinant $g$ and $\nabla_\mu$ is built from the Christoffel connection associated to $g_{\mu\nu}$. In addition, $X^\mu(\sigma)$ are a set of scalar mapping functions that determine the location of $\mathbb{M}_{p+1}$ in the spacetime $\mathcal{M}_d=\mathbb{M}_{p+1}\times\mathbb{T}^{n+2}$, while $\gamma_{ab}=g_{\mu\nu}\partial_a X^\mu\partial_b X^\nu$ is the induced metric on $\mathbb{M}_{p+1}$ and $\gamma$ its determinant while $a,b$ indices run over $\mathbb{M}_{p+1}$ (see supplementary material). The dots in \eqref{eq:stress} stand for additional quadrupole or higher corrections to the stress tensor which we will not deal with here.

The stress tensor \eqref{eq:stress} can be extracted from the metric of a black Dp brane by expanding it in powers of $\u$, in particular $g_{\mu\nu}=\eta_{\mu\nu} + h^{(M)}_{\mu\nu}+h^{(D)}_{\mu\nu}+\mathcal{O}(\u^{n+2})$ where $h^{(M)}_{\mu\nu}$ stands for the monopole contribution to the metric of order $\u^n$  and $h^{(D)}_{\mu\nu}$ for the dipole contribution of order $\u^{n+1}$. Contrary to cases in which dipole terms in \eqref{eq:stress} are viewed as perturbations, here we are interested in geometries in which monopole and dipole terms in \eqref{eq:stress} are taken to be of equal order. We can achieve this by rescaling $\u\to\lambda \u$ as well as $h^{(D)}_{\mu\nu}\to h^{(D)}_{\mu\nu}/\lambda$, and approaching the asymptotic region by sending $\lambda\to0$ while keeping $\u$ fixed. We then extract the monopole and dipole contributions to the stress tensor using linearised gravity
\begin{equation} \label{eq:stressfrommetric}
\nabla^2\bar h_{ab}^{(M)}=-16\pi G T_{ab}\hat \delta (r)~,~\nabla^2\bar h_{ai}^{(D)}=8\pi G{\mathcal{S}_{ai}}^{j}\partial_j \hat\delta(r)~~,
\end{equation}
where we have introduced $\bar h_{\mu\nu}=h_{\mu\nu}-h\eta_{\mu\nu}/2$ with $h=\eta^{\mu\nu}h_{\mu\nu}$,  and focused on the non-trivial spin components of $T^{aij}=\mathcal{S}^{aij}$ where $\mathcal{S}^{aij}=\mathcal{S}^{a[ij]}$ is the spin current and $i,j$ indices run on $\mathbb{T}^{n+2}$. Adopting the conventions of \cite{Harmark_2000} for the metric of a (boosted) spinning D3 brane, using \eqref{eq:stressfrommetric} we obtain the monopole stress tensor and spin current
\begin{equation} \label{eq:stresspin}
T^{ab}=(\varepsilon+P)u^{a}u^{b}+P\gamma^{ab}~~,~~{\mathcal{S}_{a}}^{ij}=\ell J^A \epsilon_{A}^{ij}u_a~~.   
\end{equation}
Here $T^{ab}$ takes the perfect fluid form with $\varepsilon$, $P$ and $u^a$ the energy density, pressure and fluid (boost) velocity, while $J^A$ is the angular momentum density on the Cartan plane $A=1,2,3$, $\epsilon_{A}^{ij}$ the Levi-Civita symbol on the plane $A$ and $\ell=1/\lambda$. 
The thermodynamic quantities $\varepsilon, P, J^A$ can be parametrised in terms of the horizon size $r_0$, rotation parameter $l_A$ on the plane $A$, and constant D3 brane charge $Q_3$, or equivalently in terms of the brane temperature $T$, angular velocity $\Omega_A$ and D3 charge $Q_3$ (see supplementary material). The charge $Q_3$ is constant and hence only labels solutions (is not a dynamical degree of freedom \cite{Emparan:2011hg}). Given the form of \eqref{eq:stress} and \eqref{eq:stresspin} we can find an effective theory for the deformations of spinning branes.

\vspace{0.5em}
\noindent
\textit{Effective theory and equations of motion---}
We are interested in describing the long-wavelength fluctuations of spinning Dp branes. This can be done by extending the blackfold approach \cite{Emparan:2009cs, Emparan:2009at, Emparan:2011hg, Armas:2016mes} to (strongly) spinning branes in which perturbations of spinning Dp branes are accounted for in a gradient expansion. As discussed in \cite{Armas:2016mes}, at ideal order in this gradient expansion, part of type IIB supergravity equations yield a constraint equation which is simply the conservation of the stress tensor $\nabla_\mu \bold T^{\mu\nu}=0$ in the absence of background fluxes. By integrating this equation using a Gaussian pillbox \cite{Armas:2011uf, Armas:2013hsa, Armas:2017pvj} 
one arrives at a set of equations describing the worldvolume effective theory
\begin{equation} \label{eq:eom}
\nabla_a {T^{a}}_b={\mathcal{S}^{a}}_{ij}{\Omega_{ba}}^{ij}~~,~~\tilde \nabla_a {\mathcal{S}^{a}}_{ij}=0~~,   
\end{equation}
where $\nabla_a$ is a covariant derivative compatible with both $g_{\mu\nu}$ and $\gamma_{ab}$, while $\tilde \nabla_a$ also acts on transverse $i$ indices according to $\tilde \nabla_a V^i=\partial_a V^i+{\omega_a}^{ij}V_j$ for some arbitrary vector $V^i$ and where we introduced the spin connection ${\omega_a}^{ij}=n^i_\mu \nabla_a n^{j\mu}$ with $n^\mu_j$ a set of normal vectors to $\mathbb{M}_{p+1}$ normalised such that $n^\mu_i n_{\mu}^{ j}=\delta^j_i$. In turn ${\Omega_{ab}}^{ij}$ is the field strength (or outer curvature) associated with ${\omega_a}^{ij}$, namely ${\Omega_{ab}}^{ij}=\tilde \nabla_a {\omega_b}^{ij} - \tilde \nabla_b {\omega_a}^{ij}+2{\omega_{[b}}^{ik}{\omega_{a]k}}^j$ (see supplementary material). Eqs. \eqref{eq:eom} are the equations of \emph{transverse} spin hydrodynamics. The first equation in \eqref{eq:eom} describes the non-conservation of the monopole/worldvolume stress tensor $T^{ab}$ due to a spin-curvature coupling. Backgrounds that are rotating in $\mathbb{T}^{n+2}$ have a non-trivial outer curvature, thereby leading to a non-vanishing coupling with the spin current. The second equation in  \eqref{eq:eom} states that the spin current is conserved ensuring that transverse angular momentum charges are well defined \footnote{There is also another equation that in general arises from $\nabla_\mu \bold T^{\mu\nu}=0$ describing the transverse dynamics of the brane but it is not relevant for the purposes of this letter so we present it elsewhere.}. Eqs. \eqref{eq:eom} can be written in the more common form of spin hydrodynamics as in \cite{Gallegos:2021bzp, Hongo:2021ona, Gallegos:2022jow}, which we show in the supplementary material \footnote{We note that in other formulations of hydrodynamics of spin \cite{Gallegos:2021bzp, Hongo:2021ona, Gallegos:2022jow} there is a Belinfante-Rosenfeld ambiguity in the definition of stress tensor and spin current which does not appear in the formulation presented here because the stress tensor $T^{ab}$ is symmetric.}.

The effective theory \eqref{eq:eom} can be derived from an action principle on $\mathbb{M}_{p+1}$ where the stress tensor and spin current are defined via the variations
\begin{equation}\label{eq:varS}
\delta S=\int_{\mathbb{M}_{p+1}} d^{p+1}\sigma \sqrt{-\gamma}\left(\frac{1}{2}T^{ab}\delta \gamma_{ab}+{\mathcal{S}^{a}}_{ij}\delta {\omega_a}^{ij}\right)~~.
\end{equation}
Tangential diffeormorphism transformations along $\mathbb{M}_{p+1}$ give rise, using \eqref{eq:varS}, to the first equation in \eqref{eq:eom}, while $SO(n+2)$ rotations of the normal vectors $\delta n^\mu_i={M_i}^jn^\mu_j$ for some anti-symmetric matrix $M^{ij}$ lead to the spin convservation equation in \eqref{eq:eom} since $\delta {\omega_a}^{ij}=-\tilde \nabla_a M^{ij}$. We mention that the effective theory \eqref{eq:varS} can be seen as a non-abelian gauge theory for the spin connection ${\omega_a}^{ij}$. As we are formulating a gradient expansion, we must assign a gradient ordering to each operator and source in the theory. Here we are considering the situation in which the monopole stress tensor and the spin current are equally important, thus $T^{ab},\gamma_{ab}\sim\mathcal{O}(1)$ and since ${\omega_a}^{ij}\sim\mathcal{O}(\partial)$ we must require ${\mathcal{S}^{a}}_{ij}\sim\mathcal{O}(\partial^{-1})$. This latter condition ensures that, contrary to \cite{Armas:2011uf, Armas:2013hsa, Armas:2013goa, Armas:2014rva, Armas:2017myl} where dipole terms were taken to be corrections, both monopole and dipole terms in $ \bold T^{\mu\nu}$ contribute equally at ideal order. Given the spin current in \eqref{eq:stresspin} this implies that $\ell\sim\mathcal{O}(\partial^{-1})$ and hence the parameter $\ell$ can be used as a bookkeeping parameter as in \cite{Armas:2021vku, Armas:2023tyx} for organising the gradient expansion. We will now show that the form of the monopole stress tensor and spin current for spinning D3 branes in \eqref{eq:stresspin} is a special case of a more general theory of spin hydrodynamics described by eqs.~\eqref{eq:eom}.

\vspace{0.5em}
\noindent
\textit{Degrees of freedom and conserved charges---} The starting point of any hydrodynamic theory, whenever possible, is to construct the equilibrium partition function \cite{Jensen:2012jh, Banerjee:2012iz, Armas:2013hsa, Jensen:2013kka} from which the equilibrium currents can be derived. This can be done by looking at the hydrostatic requirements for an action whose variation is \eqref{eq:varS}, in particular
\begin{equation} \label{eq:equilibriumeq}
\begin{split}
 \delta_K \gamma_{ab}&=\pounds_K \gamma_{ab}=0~~,\\
 \delta_K {\omega_a}^{ij}&=\pounds_K {\omega_a}^{ij} -\tilde\nabla_a \Lambda_K^{ij}=0~~,  
\end{split}
\end{equation}
where $K^a$ and $\Lambda_K^{ij}$ are equilibrium parameters, and $\pounds_K$ denotes the Lie derivative along $K^a$. The invariance of the conditions \eqref{eq:equilibriumeq} under tangential diffeomorphisms $\xi^a$ and rotations $M^{ij}$ requires the transformation properties for the equilibrium parameters $\delta_B K^a=\pounds_\xi K^a$ and $\delta_B \Lambda^{ij}_K=\xi^a\tilde\nabla_a\Lambda^{ij}_K-K^a\tilde\nabla_a M^{ij}$ where $B=(\xi^a,M^{ij})$. This allows us to introduce the thermal twist vector $\beta^a=u^a/T\sim\mathcal{O}(1)$ where $u^a=K^a/|K|$ is the unit normalised fluid velocity $u_a u^a=-1$ with $|K|$ the modulus of the Killing vector $K^a$ and $T=T_0/|K|$, interpreted as the temperature of the fluid, where $T_0$ is a constant global temperature. It may appear that no other ideal order scalar can be built out of the geometric data since ${\omega_a}^{ij}\sim\mathcal{O}(\partial)$ and hence also $\Lambda^{ij}_K\sim\mathcal{O}(\partial)$. However, using the bookkeeping parameter $\ell\sim\mathcal{O}(\partial^{-1})$ we can introduce the spin chemical potential $\mu^{ij}$ according to
\begin{equation} \label{eq:spinchemicalpotential}
\frac{\mu^{ij}}{T}=\ell\left(-\Lambda^{ij}_K+\beta^a {\omega_a}^{ij}\right)~~, 
\end{equation}
which is anti-symmetric in the $i,j$ indices and $\mu^{ij}\sim\mathcal{O}(1)$ as desired. The temperature and spin chemical potential constitute the basic ingredients of the equilibrium partition function
\begin{equation}\label{eq:ActionconstantT}
S_{\text{eq}}=\int_{\mathbb{M}_{p+1}} d^{p+1}\sigma \sqrt{-\gamma} P(T,\mu^{ij})~~,
\end{equation}
with $P$ the fluid pressure, and whose invariance under tangential diffeomorphisms and normal rotations yields \eqref{eq:eom}. Using variational formulae provided in the supplementary material we find the equilibrium stress tensor as in \eqref{eq:stress} together with the thermodynamic identities
\begin{equation} \label{eq:thermo}
\varepsilon+P=Ts + s_{ij}\mu^{ij}~~,~~dP=sdT+s_{ij}d\mu^{ij}~~,   
\end{equation}
where $s=\partial P/\partial T$ is the entropy density and $s_{ij}=\partial P/\partial \mu^{ij}$ the spin density, as well as the spin current
\begin{equation}
{\mathcal{S}^{a}}_{ij}=\ell s_{ij}u^a~~,    
\end{equation}
which satisfies ${\mathcal{S}^{a}}_{ij}\sim\mathcal{O}(\partial^{-1})$ by construction. Comparison with \eqref{eq:stress} one identifies $s^{ij}=J^A \epsilon_{A}^{ij}$ for the spinning D3 brane. The thermodynamic relations \eqref{eq:thermo} also match those of the spinning D3 brane (see supplementary material) which confirm that the gradient ordering of \eqref{eq:spinchemicalpotential} is appropriate since the angular momenta of spinning Dp branes is not a perturbative quantity in gravity. The common lore of hydrodynamics is to promote the equilibrium parameters $K^a$ and $\Lambda^{ij}_K$ to slowly varying functions out of equilibrium, allowing us to identify $T,\mu^{ij}, u^a$ as the dynamical degrees of freedom of spin hydrodynamics whose gradients correct the currents $T^{ab}$ and ${\mathcal{S}^{a}}_{ij}$.

 In equilibrium, eqs.~\eqref{eq:eom} have two sets of conserved charges that can be obtained by integrating the conserved currents $T^a=T^{ab}k_b+{\mathcal{S}^{a}}_{ij}n^{i\mu}n^{j\nu}\nabla_\nu k_\mu$ and ${\mathcal{S}^{a}}_{ij}$ over a spatial slice of $\mathbb{M}_{p+1}$. The current $T^a$ gives rise to a total conserved energy if the Killing vector $k^\mu=k^a \partial_a X^\mu$ is the generator of time translations, and to linear/angular momentum in $\mathbb{M}_{p+1}$ when $k_a$ is the generator of spatial translations/rotations. In turn, the current ${\mathcal{S}^{a}}_{ij}$ leads to a set of spin charges $S^{ij}$. In the specific case in which the configuration is rigidly rotating in $\mathbb{T}^{n+2}$, as for spinning D3 branes, there is an associated set of transverse Killing vector fields $\zeta_{\mu}^{ij}$ for each plane $(i,j)$. The conserved current associated with $\zeta_{\mu}^{ij}$ can be obtained using the full spacetime energy-momentum tensor \eqref{eq:stress} by integrating the conserved current $\bold T^{\mu\nu}\zeta_{\nu}^{ij}$ over a spatial slice \cite{Armas:2013hsa, Armas:2014rva, Armas:2017pvj}. In this case, for which transverse spacelike vector fields are available, the spin charges $S^{ij}$ have the physical interpretation of (transverse) angular momenta in $\mathbb{T}^{n+2}$.

\vspace{0.5em}
\noindent
\textit{Spin current as the holographic dual of the R-current---}
The $\mathcal{N}=4$ SYM theory has a global $SO(6)$ R-symmetry which rotates the supercharges one into another. AdS/CFT geometrizes this symmetry in 10 dimensions by mapping in the decoupling limit the charges of the R-symmetry group to angular momenta of the D3 brane in the
planes perpendicular to its worldvolume. As mentioned, the current whose integral generically gives the (transverse) angular momenta is the spin current ${\mathcal{S}^{a}}_{ij}$. This clearly implies a holographic relation between the (near-horizon) spin current of the spinning D3 brane and the 4d R-current of strongly coupled $\mathcal{N}=4$ SYM. For instance, the transverse indices $(ij)$ in the spin current are dual to the indices carried by the SO(6) charges.

The aforementioned holographic relation can immediately be made precise once we specialize to thermal states whose gravity duals are spinning D3 branes with equal angular momenta in the 3 Cartan planes of rotation. Such solutions fit into a Kaluza-Klein (KK) ansatz which when inserted into the Type IIB supergravity equations of motion leads to a Einstein-Maxwell-Chern-Simons 5d theory (see for instance \cite{Cvetic__1999}). With a suitable choice of co-ordinates, the spin current is contained in the 10d metric component mixing transverse angles with worldvolume directions. In the large $r$ or equivalently near-boundary expansion, the spin current is extracted from the near-horizon geometry as the coefficient of the term scaling as $\sim \mu_{i}^2 d\phi_{i}/r^2$. Upon the above KK reduction, this coefficient precisely gives rise to the (3 copies of) gauge-covariant 4d R-current(s) $\mathcal{J}_{a}^{A}$, 
where the index $A=1,2,3$ labels the 3 copies of the R-current and it is left there for clarity. We can then identify the near-horizon spin current ${\mathcal{S}^{a}}_{ij}$ as the dual of the 4d R-current of strongly coupled $\mathcal{N}=4$ SYM in the equally spinning case via ${\mathcal{S}_{a}}^{ij}\to \ell \mathcal{J}_{a}^{A}$. Note that for this natural identification to be possible, a treatment of spinning D-branes in the strongly spinning regime is essential.

\vspace{0.5em}
\noindent
\textit{Anomalous spin hydrodynamics---}
The duality between spin current and R-current implies that the former must also be an anomalous current. The effective theory governing the long wavelength dynamics of black branes is a perturbative expansion in gradients truncated at a given order. Corrections to the effective theory at this given order can arise due to three different sources as explained in \cite{Armas:2011uf, Armas:2016mes}: hydrodynamic corrections to the currents $T^{\mu\nu}$ and $T^{\mu\nu\rho}$, multipole corrections to \eqref{eq:stress}, and corrections either due to modifications of the asymptotic structure or the presence of non-trivial fluxes that lead to modifications of the conservation law for the stress tensor, that is $\nabla_\mu \bold T^{\mu\nu}=\bold f^{\nu}$ where $\bold f^{\mu}$ is a forcing function that admits a multipole expansion as in \eqref{eq:stress} (see supplementary material). Here we focus on the decoupling limit of the D3 brane within the KK ansatz of \cite{Cvetic__1999} for which both hydrodynamic corrections and corrections due to the presence of fluxes arise, in particular $\bold f^{\mu}=F^{\mu\nu_1...\nu_4}\bold{J}_{\nu_1...\nu_4}/4!$ where $F^{\mu\nu_1...\nu_4}$ is the five-form flux and $\bold{J}_{\nu_1...\nu_4}$ the D3-brane current \cite{Armas:2016mes}. In this limit, the presence of fluxes leads to anomalous contributions to the spin conservation equation \eqref{eq:eom} \cite{Son:2009tf, Erdmenger:2014jba}, such that from a 10d perspective 
\begin{equation} \label{eq:anomalousspin}
\tilde \nabla_a {\mathcal{S}^{a}}_{ij}=\frac{\ell^3}{8} C_{ijklmn}\epsilon^{abcd}{\Omega_{ab}}^{kl}{\Omega_{cd}}^{mn}~~,    
\end{equation}
where $C_{ijklmn}$ is a constant matrix of anomaly coefficients, anti-symmetric in each pair $(i,j),(k,l),(m,n)$ and symmetric under exchange of pairs, while $\epsilon^{abcd}$ is the Levi-Civita tensor in $\mathbb{M}_4$ and where $\ell^3$ guarantees the correct order in the expansion. The relation between the anomaly term and the forcing function $\bold f^{\mu}$ is given in the supplementary material. We note that the five-form flux does not appear explicitly in \eqref{eq:anomalousspin} because the KK ansatz \cite{Cvetic__1999} for the flux is determined in terms of 10d metric components and its derivatives, effectively generating the anomaly. As with the usual chiral anomaly, the anomalous term in the conservation law \eqref{eq:anomalousspin} is generated using anomaly inflow (see e.g. \cite{Jensen:2012kj, Jensen:2013kka}) in which the currents in \eqref{eq:varS} should now be viewed as covariant currents acquiring modifications due to inflow while the partition function \eqref{eq:ActionconstantT} receives non-gauge invariant corrections (see e.g. \cite{Ammon:2020rvg}).

The form of the RHS of \eqref{eq:anomalousspin}, modifying the conservation law for $\bold T^{\mu\nu}$, makes it clear that it is a gravitational anomaly. Not only is ${\Omega_{ab}}^{kl}$ purely geometric, it is also given in terms of components of the Riemann tensor (see supplementary material). This interpretation is also manifest by the nature of the conserved charges. 

Given the structure of the equation \eqref{eq:anomalousspin} at first order, it is straightforward to find the corrected constitutive equations by requiring the second law of thermodynamics $\nabla_a S^a\ge0$, where $S^a=P\beta^a-T^{ab}\beta_b-\ell^{-1}\mu^{ij}{\mathcal{S}^{a}}_{ij}/T+S_{\text{nc}}^a$ is the entropy current and $S_{\text{nc}}^a$ the non-canonical contribution to the entropy current. Parametrizing the corrections to the currents as $T^{ab}=(\varepsilon+P)u^{a}u^{b}+P\gamma^{ab}+\mathcal{T}^{ab}$ and ${\mathcal{S}^{a}}_{ij}=\ell s_{ij}u^a+{\Sigma^a}_{ij}$, and using the first equation in \eqref{eq:eom} as well as \eqref{eq:anomalousspin} we can determine the corrections to the currents in the Landau frame ($u_b\mathcal{T}^{ab}=u_a {\Sigma^a}_{ij}=0$) to be
\begin{equation} \label{eq:correctedcurrents}
\begin{split}
\mathcal{T}^{ab}=&-\eta \sigma^{ab}-\zeta \theta P^{ab}+\mathcal{O}(\partial^2)  \\
{\Sigma^a}_{ij}=&-\ell \mathcal{D}_{ijkl}P^{ab}\left(\ell \beta^c{\Omega_{cb}}^{kl}+\tilde\nabla_b\left(\frac{\mu^{kl}}{T}\right)\right)\\
&+\ell {\xi_{ij}}{\varpi^{a}}+\ell^2{\xi^S_{ijkl}}{B^{akl}}+\mathcal{O}(\partial)~~,
\end{split}
\end{equation}
where $\eta,\zeta,\mathcal{D}_{ijkl}\ge0$ are the shear viscosity, bulk viscosity and spin diffusion coefficients respectively. We have also introduced the fluid shear tensor $\sigma^{ab}=P^{ac}P^{bd}(2\nabla_{(c}u_{d)}-P_{cd}\theta/p)$ and expansion $\theta=\nabla_a u^a$ with $P^{ab}=u^{a}u^{b}+\gamma^{ab}$ the projector orthogonal to $u_a$. In turn ${\xi_{ij}}$ is an anti-symmetric matrix of coefficients yielding a non-zero chiral vortical effect where $\varpi^a=\epsilon^{abcd}u_b\nabla_cu_d/2$ is the fluid vorticity, while ${\xi^S_{ijkl}}$ is a matrix of coefficients anti-symmetric in the pairs $(i,j)$ and $(k,l)$ responsible for a chiral spin effect (the analogue of the chiral magnetic effect), where we have introduced the "magnetic" outer curvature $B^{akl}=\epsilon^{abcd}u_b{\Omega_{cd}}^{kl}/2$. An analogous computation to \cite{Son:2009tf, Neiman:2010zi}, sets
\begin{equation}
 {\xi_{ij}}=\frac{\partial \tilde \xi}{\partial (\mu^{ij}/T)}|_P ~~,~~ \xi^S_{ijkl} =\frac{\partial \tilde \xi^S_{kl}}{\partial (\mu^{ij}/T)} |_P~~,
\end{equation}
where $\tilde \xi$ and $\tilde \xi^S_{kl}$ are the coefficients appearing in $S^a_{\text{nc}}=\tilde \xi\varpi^a+\ell \tilde \xi^S_{kl} B^{akl}+\mathcal{O}(\partial^2)$ and given by $\tilde \xi=C_{ijklmn}\mu^{ij}\mu^{kl}\mu^{mn}/3T$ and $\tilde \xi^S_{ij}=C_{ijklmn}\mu^{kl}\mu^{mn}/2T$, where we have neglected other potential contributions to $\tilde \xi$ and $\tilde \xi^S_{ij}$ \cite{Neiman:2010zi} that are not relevant for the D3 brane. Thus, in this case, $\tilde \xi$ and $\tilde \xi^S_{ij}$ are exclusively determined in terms of the anomaly coefficients $C_{ijklmn}$. 

This analysis matches precisely the hydrodynamics of R-charged black holes when the spin current is holographically identified as the R-current. For instance, it is now a simple task to translate the conservation laws defining the 10-dimensional anomalous hydrodynamics with transverse spin to the holographic Ward identities of a 4d theory exhibiting a chiral anomaly. Turning on a background field strength in the 4d anomalous theory means in 10d supergravity turning on a background outer curvature. Starting from the first equation in \eqref{eq:eom} and \eqref{eq:anomalousspin} in the equally spinning case, and using the holographic relations ${\mathcal{S}_{a}}^{ij}\to \ell \mathcal{J}_{a}^{A}$, $2\ell {\Omega_{ab}}^{ij}\to F_{ab}^A$ where $F_{ab}^{A}$ are the field strengths of the 4d theory, we arrive at the standard conservation laws of a theory with a chiral anomaly \cite{Son:2009tf}. 

\vspace{0.5em}
\noindent
\textit{Discussion---}
Using spinning D3 branes as a guiding example, we have given a geometric interpretation of the chiral anomaly, in particular the chiral anomaly in 4d is seen as a gravitational anomaly from a 10d point of view. In the low energy hydrodynamic regime of QFT with a chiral anomaly, the dynamics of the boundary stress tensor and R-current of anomalous R-charged fluids in 4d is mapped to the dynamics of the spacetime stress tensor \eqref{eq:stress} of anomalous spinning fluids in 10d. In the context of $\mathcal{N}=4$ SYM, we showed that the anomalous R-current in 4d holography corresponds to an anomalous spin current in 10d. This point of view is valuable for studying instabilities of Dp-branes. For instance, using the exact coefficients for the spin current \eqref{eq:correctedcurrents} of the D3 brane given in the supplementary material, it is straightforward to perform a linearized analysis and find, directly in 10d, the hydrodynamic instability in the spin diffusion modes recently uncovered in \cite{Gladden:2024ssb} from a 5d perspective. More broadly, we uncovered a connection between two seemingly different aspects of the quark-gluon plasma: the chiral anomaly and spin degrees of freedom. 

From a 10d string theory point of view, however, the low energy effective theory of spinning D3-branes is that of a spinning fluid carrying a higher-form D3-brane charge coupled to a non-trivial background five-form flux such that $\bold f^{\mu}=F^{\mu\nu_1...\nu_4}\bold{J}_{\nu_1...\nu_4}/4!$. Following the same procedure outlined around \eqref{eq:eom}, leads to the modified spin current conservation law
\begin{equation} \label{eq:anomalousspinHF}
\tilde \nabla_a {\mathcal{S}^{a}}_{ij}=\frac{1}{4!}{J^{\mu_1...\mu_4}}_{[i}F_{j]\mu_1...\mu_4}~~,    
\end{equation}
where $J^{\mu_1...\mu_4 i}$ is the dipole contribution to the current $\bold{J}_{\nu_1...\nu_4}$. In particular, the D3-brane at ideal order is characterized by a magnetic dipole moment $J^{abcij}=\ell\epsilon^{acbd}u_d s^{ij}$. We see that the coupling between dipole contributions and five-form flux \eqref{eq:anomalousspinHF} violates spin conservation, similarly to charged spinning point particles coupled to electromagnetic fields (see e.g.\cite{Costa:2012cy}), and hence can effectively generate anomalous contributions for special classes of backgrounds \footnote{This idea is similar to the one implemented in \cite{Casero:2007ae}.}. We intend to develop a hydrodynamic theory of higher-form spinning fluids along the lines of \cite{Armas:2018ibg} in order to trace the string theoretic origin of the chiral anomaly in holographic fluids.

The effective theory for anomalous spinning fluids that we introduced in this letter is only expected to describe the low energy dynamics of spinning D3 branes in the near-horizon limit. Moving beyond the near-horizon limit and for general perturbations, we expect that additional backreaction corrections will kick in that will also modify the conservation law for the monopole stress tensor $T^{ab}$ as in \cite{Caldarelli:2013aaa}. We also expect that many other transport coefficients such as the Young modulus \cite{Armas:2011uf} will begin to play a role. To understand this further one would need to study the deformations of spinning D3 branes as in \cite{Emparan:2013ila, DiDato:2015dia}, and complete the analysis of \cite{Erdmenger:2014jba} to the scalar sector together with an in-depth analysis of the uplifted solution. 

In the context of heavy-ion collisions, there are various formulations of spin hydrodynamics \cite{Huang:2024ffg, Florkowski:2024cif, Becattini:2024uha}, some of which instead have \emph{intrinsic} spin \cite{Gallegos:2021bzp, Hongo:2021ona, Gallegos:2022jow}, i.e. the spin lives on $\mathbb{M}_{p+1}$ rather than on $\mathbb{T}^{n+2}$, and other formulations that do not include spin degrees of freedom \cite{Becattini:2013fla, Florkowski:2017ruc, Becattini:2021iol}. The formulation presented here is yet another form of spin hydrodynamics but we expect various connections with theories with \emph{intrinsic} spin \cite{Gallegos:2021bzp, Hongo:2021ona, Gallegos:2022jow} which will be explored elsewhere \footnote{One point of departure with theories with \emph{intrinsic} spin is that in this formulation with \emph{transverse} spin, coupling to the sources $\gamma_{ab}$ and ${\omega_a}^{ij}$ directly in \eqref{eq:varS} and not to the tangent $\partial_a X^\mu$ and normal $n^\mu_i$ vectors, does not lead to a relation between the spin chemical potential $\mu^{ij}$ and the thermal vorticity $\partial_{[a}u_{b]}$ of the fluid in equilibrium.}. To note is the relation between the R-current of $\mathcal{N}=1$ theories and the \emph{intrinsic} spin current \cite{Cartwright:2024dcj}. Using the construction of NS5-D3 branes in Polchinski-Strassler as in \cite{Armas:2022bkh}, it would be relevant to understand if the spin current discussed in this letter is related to the R-current in $\mathcal{N}=1$. It would be interesting to see if, for all spinning Dp branes, the \emph{transverse} spin current is directly related to the \emph{intrinsic} spin current, thus providing an easy method for extracting transport properties of spin hydrodynamics from holography. 


\vspace{0.5em}
\begin{acknowledgments}
\noindent
\textit{Acknowledgments---}%
We would like to acknowledge fruitful discussions with, or comments on an earlier draft from, D. Berenstein, J. Bhattacharya, J. de Boer, U. G\"{u}rsoy, T. Harmark, A. Jain, P. Kovtun, K. Landsteiner, N. Obers, K. Schalm, R. Singh, C. Vafa and A. Zhiboedov. JA is partly supported by the Dutch Institute for Emergent Phenomena (DIEP) cluster at the University of Amsterdam via the DIEP programme Foundations and Applications of Emergence (FAEME). This work is dedicated to the memory of Umut G\"{u}rsoy who worked on the two topics of this letter and with whom we had enlightening discussions during the completion of this work.
\end{acknowledgments}

\bibliography{bibliography}

\clearpage

\appendix 

\onecolumngrid

\renewcommand{\thesection}{\Alph{section}}
\renewcommand{\thesubsection}{\thesection.\arabic{subsection}}
\renewcommand{\thesubsubsection}{\thesubsection.\arabic{subsubsection}}
\makeatletter
\renewcommand{\p@section}{}
\renewcommand{\p@subsection}{}
\renewcommand{\p@subsubsection}{}
\makeatother

\counterwithin{figure}{section}
\renewcommand\thefigure{\Alph{section}.\arabic{figure}}
\renewcommand{\theequation}{\Alph{section}.\arabic{equation}}
\setcounter{secnumdepth}{2}
{\center{\large\bfseries Supplementary Material}\par}

\vspace{0.5cm}

\section{Geometry of embedded spaces} \label{app:geometry}
In this section we give details on the geometry of embedded spaces used in the core of the paper following \cite{Armas:2013hsa, Armas:2017pvj}. We consider d-dimensional spacetimes with a product space $\mathcal{W}_d=\mathbb{M}_{p+1}\times \mathbb{T}^{n+2}$ where $d=n+p+3$. The spacetime $\mathcal{W}_d$ has metric $g_{\mu\nu}$ and coordinates $x^\mu$ where the Greek indices run from $\mu=0,...,d-1$. We introduce the set of mapping functions $X^\mu(\sigma^a)$ such that $x^\mu=X^\mu$ specifies the location of $\mathbb{M}_{p+1}$ within $\mathcal{W}_d$ and where $\sigma^a$ are coordinates on $\mathbb{M}_{p+1}$. The Latin indices run from $a=0,...,p-1$. Given the set of mapping function $X^\mu$ we can define the tangent vectors $\mathbb{M}_{p+1}$ as $t^\mu_a=\partial_a X^\mu$ and implicitly the normal vectors $n^\mu_i$ such that
\begin{equation}
g_{\mu\nu}t^\mu_a t^\nu_b=\gamma_{ab}~~,~~g_{\mu\nu}t^\mu_a n^\nu_i=0~~,~~g_{\mu\nu}n^\mu_i n^\nu_j=\delta_{ij}~~,
\end{equation}
where the first condition is the definition of the induced metric on $\mathbb{M}_{p+1}$, the second the orthogonality condition of the normal vectors, and in the third we used diffeomorphisms to set the normalization of the normal vectors equal to $\delta_{ij}$. This latter choice leaves a residual $SO(n+2)$ rotation symmetry of the normal vectors 
\begin{equation} \label{eq:normal}
n^\mu_i\to {\mathcal{M}_i}^{j}n^\mu_j~~,   
\end{equation}
where ${\mathcal{M}_i}^{j}$ is a matrix in $SO(n+2)$. Infinitesimally we have that $\delta n^\mu_i= {M_i}^{j}n^\mu_j$ with ${M_i}^{j}$ an anti-symmetric matrix in $i,j$. The transverse Latin indices run from $i=1,...,n+2$. To further specify the product space we introduce the Gauss-Weingarten equations
\begin{equation}
\begin{split}
t^{\mu}_a \nabla_\mu t^\nu_b=\gamma_{ab}^c t_c^\nu+K_{ab}^i n_i^\nu~~,~~t^\mu_a \nabla_\mu n^{\nu i}= -K_{ab}^i t^{b\nu}-{\omega_a}^{ij}n_j^\nu~~,
\end{split}
\end{equation}
with $\nabla_\mu$ the covariant derivative compatible with $g_{\mu\nu}$, $\gamma_{ab}^c$ the affine connection built from the induced metric $\gamma_{ab}$, $K_{ab}^i=n^i_\mu \nabla_a t_b^\mu=K_{ba}^i$ the extrinsic curvature and ${\omega_a}^{ij}=n^i_\mu\nabla_a n^{j\mu}=-{\omega_a}^{ji}$ the spin connection. Here we have also introduced the covariant derivative $\nabla_a$ compatible with both $g_{\mu\nu}$ and $\gamma_{ab}$ acting on an arbitrary vector $v^\mu_a$ as $\nabla_b v^\mu_a=\partial_b v^\mu_a-\gamma_{ba}^{c}v^\mu_c+\Gamma^{\mu}_{\nu\lambda}v^\nu_a t^\lambda_b$ where $\Gamma^{\mu}_{\nu\lambda}$ is the affine connection built using $g_{\mu\nu}$. Under the infinitesimal version of \eqref{eq:normal} the spin connection transforms as $\delta {\omega_a}^{ij}=-\tilde \nabla_a M^{ij}$, where $\tilde \nabla_a$ acts on arbitrary vectors $v^\mu_{ai}$ as $\tilde \nabla_b v^\mu_{ai}=\partial_b v^\mu_{ai}-\gamma_{ba}^{c}v^\mu_{ci}+\Gamma^{\mu}_{\nu\lambda}v^\nu_{ai} t^\lambda_b+{\omega_{bi}}^{j}v_{aj}^\mu$. Given the transformations under normal rotations it is useful to define the outer curvature tensor as
\begin{equation}
{\Omega_{ab}}^{ij}=\tilde \nabla_a {\omega_b}^{ij} - \tilde \nabla_b {\omega_a}^{ij}+2{\omega_{[b}}^{ik}{\omega_{a]k}}^j~~,
\end{equation}
which transforms covariantly under rotations and satisfies the identity $\tilde \nabla_{[c}{\Omega_{ab]}}^{ij}=0$. The outer curvature is related to the extrinsic curvature and Riemann tensor due to the Ricci-Voss integrability condition
\begin{equation} \label{eq:RicciVoss}
{R_{ab}}^{ij}={\Omega_{ab}}^{ij}-2K_{c[a}^i{K_{b]}}^{cj}~~,
\end{equation}
where ${R_{ab}}^{ij}=t_a^\mu t_b^\nu n^{\lambda i}n^{\sigma j} R_{\mu\nu\lambda\sigma}$ and $R_{\mu\nu\lambda\sigma}$ is the Riemann tensor associated with $g_{\mu\nu}$. In the case in which $\mathbb{T}^{n+2}$ has $SO(n+2)$ symmetry there exists a set of transverse Killing vector fields $\zeta_\mu^{ij}$, one for each Cartan plane, such that $t^\mu_a \zeta_\mu^{ij}=0$. In this case, for which the spinning D3 brane is an example, one can show that $K_{ab}^k \zeta_k^{ij}=K_{ab}^j {{\omega_a}^i}_j=0$ \cite{Armas:2014rva}. By extension ${\mathcal{S}_a}^{ij}K_{bcj}=0$ which follows due to the decomposition ${\omega_{a}}^{ij}=\sum_{A}{\omega_{a}}^{A}\epsilon_A^{ij}$ where $\omega_a^A=\epsilon^{A}_{ij}{\omega_{a}}^{ij}/2$ and $A$ is a label for each Cartan plane characterised by a given pair of $(i,j)$ indices while $\epsilon_A^{ij}$ the Levi-Civitta symbol on the plane $A$ \cite{Armas:2014rva}. In turn this implies that the spin anomaly can be written as
\begin{equation} \label{eq:anomalousspin1}
\begin{split}
\tilde \nabla_a {\mathcal{S}^{a}}_{ij}=\frac{\ell^3}{8}C_{ijklmn}\epsilon^{abcd}{\Omega_{ab}}^{kl}{\Omega_{cd}}^{mn}=\frac{\ell^3}{8}C_{ijklmn}\epsilon^{abcd}{R_{ab}}^{kl}{R_{cd}}^{mn}~~,    
\end{split}
\end{equation}
where we used the Ricci-Voss identity \eqref{eq:RicciVoss}. This highlights the fact that we are dealing with a gravitational anomaly.

\vspace{0.5em}
\noindent
\textit{Geometric variations---} In the letter we made use of variational formulae for making various manipulations. In particular, in order to derive the spin hydrodynamic equations \eqref{eq:eom} we used the tangential variations of the induced metric $\delta_{\xi} \gamma_{ab}=2\nabla_{(a}\xi_{b)}$
which is obtained by performing a tangential diffeomorphism $x^\mu\to x^\mu-\xi^\mu$ where $\xi^\mu=t^\mu_a \xi^a$ for an arbitrary vector $\xi^a$, as well as tangential variations of the spin connection
\begin{equation}
\begin{split}
\delta_{\xi}{\omega_a}^{ij}=&\xi^b\nabla_b {\omega_a}^{ij}+{\omega_b}^{ij}\nabla_a\xi^b~~. 
\end{split}
\end{equation}
In turn, in order to derive the equilibrium currents from an equilibrium partition function we used the variational formulae under infinitesimal tangential diffeomorphisms and rotations with parameters $B=(\xi^a,M^{ij})$ such that
\begin{equation}
\delta_B\gamma_{ab}=2\nabla_{(a}\beta_{b)}~~,~~\ell\delta_B {\omega_{a}}^{ij}=\ell\beta^b {\Omega_{ba}}^{ij}+\tilde\nabla_a\left(\frac{\mu^{ij}}{T}\right)~~,
\end{equation}
where $\beta_a=u_a/T$ with $u^{a}$ the fluid velocity, $T$ the fluid temperature, and $\mu^{ij}$ the spin chemical potential. We also used that the variations of the temperature and spin chemical potential take the form $\delta_B T=\frac{T}{2}u^{a}u^{b}\delta_B \gamma_{ab}$ and $
\delta_B \mu^{ij}=\frac{\mu^{ij}}{2}u^{a}u^{b}\delta_B\gamma_{ab}+T \beta^a\delta_B{\omega_{a}}^{ij}$, respectively.

\vspace{0.5em}
\noindent
\textit{Spin hydrodynamic equations---} The conservation laws for the monopole stress tensor and spin current in \eqref{eq:eom} can be written in the more standard form of spin hydrodynamics found in \cite{Gallegos:2021bzp, Hongo:2021ona, Gallegos:2022jow}. For this purpose we introduce the covariant derivative $\bar \nabla_\mu =\gamma^{\nu}_\mu \nabla_\nu$ with $\gamma_{\mu\nu}=t_\mu^a t_\nu^b \gamma_{ab}$ being the tangential spacetime projector onto $\mathbb{M}_{p+1}$. In this case we rewrite the spin hydrodynamic equations as 
\begin{equation} \label{eq:spacetime}
\gamma^\sigma_\alpha \bar \nabla_\lambda\bar{\mathcal{T}}^{\lambda\alpha}=\gamma^\sigma_\alpha{\mathcal{S}^{\mu\lambda\rho}}{R^\alpha}_{\mu\lambda\rho}~~,~~\bar \nabla_\mu \mathcal{S}^{\mu\nu\rho}=2 \bar{\mathcal{T}}^{[\nu\rho]}~~,
\end{equation}
where we have defined the effective stress tensor $\bar{\mathcal{T}}^{\mu\nu}=T^{\mu\nu}-2{{\mathcal{S}^{\lambda}}_{\alpha}}^\nu {K_\lambda}^{\mu\alpha}$, which is not in general symmetric in the $\mu,\nu$ indices, and we have used the tangential and normal vectors to define, for instance, ${\mathcal{S}^{\mu\lambda\rho}}=t^{\mu a}n_i^\lambda n^\rho_j {\mathcal{S}_{a}}^{ij}$ and similarly for the other tensors. The form of \eqref{eq:spacetime} is precisely the form found in \cite{Gallegos:2021bzp, Hongo:2021ona, Gallegos:2022jow}.
\vspace{0.5em}

\textit{D3-brane equations coupled to the five-form flux---} Here we describe how to model corrections due to the presence of fluxes.  The spinning D3-brane couples to the background metric $g_{\mu\nu}$ and the five-form field strength $F^{\mu\nu\lambda\rho\sigma}$ as in \cite{Armas:2016mes} leading to the modified conservation law
\begin{equation}\label{eq:fullstc}
\nabla_\mu \boldsymbol{T}^{\mu\nu}=F^{\nu\mu\lambda\rho\sigma}\boldsymbol{J}_{\mu\lambda\rho\sigma}, 
\end{equation}
where the first term is the Lorentz force induced by the five~form, $\boldsymbol{J}_{\mu\lambda\rho\sigma}$ is the D3-brane current that admits a multipole expansion similar to \eqref{eq:stress} and obeys the conservation law $\nabla_\mu \boldsymbol{J}^{\mu\lambda\rho\sigma}=0$. Higher-form current conservation equations of the type $\nabla_\mu \boldsymbol{J}^{\mu\lambda\rho\sigma}=0$ have been considered in \cite{Armas:2012ac, Armas:2013aka} and result in trivial dynamics along the worldvolume that set the D3-brane charge constant.

Instead of considering \eqref{eq:fullstc} we can, in the spirit of \cite{Bhattacharyya:2008ji} where non-flat corrections to the geometry are introduced via a suitable forcing field, model the effect of fluxes by introducing a forcing function such that
\begin{equation} \label{eq:force}
\nabla_\mu \boldsymbol{T}^{\mu\nu}=\bold f^{\nu}~~,
\end{equation}
where the force $\bold f^{\nu}=F^{\nu\mu\lambda\rho\sigma}\boldsymbol{J}_{\mu\lambda\rho\sigma}/4!$ can be expanded in a multipole series as 
\begin{equation}
\bold f^{\mu}=f^{\mu}(\sigma)\hat \delta(X)-\nabla_\rho\left(f^{\mu\rho}(\sigma)\hat \delta (X) \right)+...~~,    
\end{equation}
and where $f^\mu$ and $f^{\mu\rho}$ are monopole and dipole contributions respectively. Integrating \eqref{eq:force} using a Gaussian pillbox as in \cite{Armas:2017pvj} we find the modified equations on $\mathbb{M}_{p+1}$, namely 
\begin{equation}
\nabla_a {T^{a}}_b={\mathcal{S}^{a}}_{ij}{\Omega_{ba}}^{ij}-\left(f_{b}-t_{b\nu}\nabla_a f^{\nu a}\right)~~,~~\tilde \nabla_a {\mathcal{S}^{a}}_{ij}=-f_{[ij]}~~.
\end{equation}
In the near-horizon limit of the spinning D3-brane we can model the effect of the anomaly by setting $f_b=f_{\nu a}=0$ and $f_{[ij]}=-\ell^3 C_{ijklmn}\epsilon^{abcd}{\Omega_{ab}}^{kl}{\Omega_{cd}}^{mn}/8$. Comparing with \eqref{eq:anomalousspinHF} we note that $f^{[ij]}=-\frac{1}{4!}J^{\mu_1...\mu_4 [i}{F^{j]}}_{\mu_1...\mu_4}$.

\section{Thermodynamics and currents of spinning D3 branes} \label{app:Dbrane}
In this section we provide the thermodynamics of spinning D3 branes and their behaviour in the near-horizon limit, making contact with the spin-hydrodynamics theory formulated in the main text. The (boosted) spinning D3 brane has a monopole stress tensor of the form $T^{ab}=(\varepsilon+P)u^{a}u^{b}+P\gamma^{ab}$ where the energy density $\varepsilon$ and pressure $P$ are given by
\begin{equation}\label{eq:D3eandp}
    \varepsilon=\frac{\Omega_{5}}{16\pi G}r_{0}^4(4\sinh^2{\alpha}+5)~,~ P=-\frac{\Omega_{5}r_{0}^4}{16\pi G}(4\sinh^2{\alpha}+1)~~,
\end{equation}
where $r_0$ is the brane thickness, $\alpha$ the charge parameter, $\Omega_5$ the volume of the 5-sphere and $G$ Newton's constant in 10 spacetime dimensions. The D3 brane is also characterised by a constant D3 brane charge $Q_3$ given by 
\begin{equation}
Q_{3}=\frac{\Omega_{5}}{16\pi G}4r_{0}^{4}\cosh{\alpha}\sinh{\alpha}~~.  
\end{equation}
The D3 brane thermodynamics also includes the temperature $T$ and entropy $s$ given by
\begin{equation}
   T=\frac{4-2\sum_{A=1}^{3}\frac{l_{A}^2}{l_{A}^2+r_{H}^2}}{4\pi r_{H} \cosh{\alpha}}~~,~~ s=\frac{\Omega_{5}}{4G}r_{0}^4 r_{H}\cosh{\alpha}~~,
\end{equation}
where $l_A$ is the angular rotation parameter on each of the Cartan planes $A=1,2,3$. In turn, $r_H$ is the horizon size, which can be obtained by evaluating the largest real root of $\bar f(r_H)=0$ where
\begin{equation}
\bar{f}(r)=1-\frac{1}{L(r)}\frac{r_{0}^{4}}{r^{4}}~~,~~L(r)=\prod_{A=1}^{3}\left(1+\frac{l_{A}^2}{r^2}\right)~~, 
\end{equation}
and whose exact expression in general is cumbersome. In addition, the D3 brane is also characterised by an angular velocity and angular momentum associated to each Cartan plane, namely
\begin{equation}
\Omega_A= \frac{l_{A}}{(l_{A}^2+r_{H}^2)\cosh{\alpha}}~~,~~J^A=\frac{\Omega_{5}}{8\pi G}r_0^4\cosh\alpha l^A~.
\end{equation}
These quantities satisfy the thermodynamic identities $\varepsilon+P=sT+\Omega_A J^A$ and $dP=sdT+J^Ad\Omega_A$, where the sum over the $A$ indices is implicit, and we further note that the charge $Q_3$ is constant and kept fixed. Using the procedure explained in the main text, we can compute the spin current of the D3 brane which takes the form
\begin{equation}
     {\mathcal{S}_{a}}^{ i j}=\ell u_a J^A \epsilon^{ij}_A~~,
\end{equation}
from which we can extract the spin density $s^{ij}=J^A \epsilon^{ij}_A$. In turn we can straightforwardly define the spin chemical potential $\mu^{ij}=\frac{1}{2}\Omega^A \epsilon^{ij}_A$ and hence we can equivalently write the Euler relation $\varepsilon+P=sT+\mu^{ij}s_{ij}$ as in the main text. All thermodynamic quantities of interest such as the pressure can be written as $P(T,\mu^{ij};Q_3)$. Here $T,\mu^{ij}$ are the true dynamical variables and since $Q_3$ is constant, it only labels solutions so we can simply omit it and write instead $P(T,\mu^{ij})$ \cite{Emparan:2011hg}. 

\vspace{0.5em}
\noindent
\textit{Near-horizon limit---}
The near-horizon limit of spinning D3 branes has been studied in \cite{Harmark_2000} and here we use the scalings introduced in \cite{Harmark_2000} to obtain and rewrite the near-horizon thermodynamics in the language of the hydrodynamic theory with spin introduced in the main text. In particular we introduce the dimensionfull parameter $\mathcal{L}$ such that the following ratios 
\begin{equation}\label{eq:rescaling}
\begin{split}
    r_{0}=\frac{r_{0,old}}{\mathcal{L}^2}\quad,\quad l_{A}=\frac{l_{A,old}}{\mathcal{L}^2}\quad,\quad \mathcal{H}^4=\frac{\mathcal{H}_{old}^4}{\mathcal{L}^4}~~,~~G=\frac{G_{old}}{\mathcal{L}^8}~,
    \end{split}
\end{equation}
are held fixed when $\mathcal{L}\to0$ and where we defined $\mathcal{H}_{old}^4=r_0^4\cosh\alpha\sinh\alpha$. In this limit the energy density and pressure in \eqref{eq:D3eandp} diverge. We can obtain a finite result for the monopole stress tensor by subtracting from it the stress tensor of the ground state as follows
\begin{equation}
T^{ab}=T^{ab}_{old}+Q_3\gamma^{ab}~~,    
\end{equation}
which leads to the finite energy density and pressure of the strongly coupled $\mathcal{N}=4$ SYM plasma, namely
\begin{equation}
    \varepsilon=\frac{3\Omega_{5}}{16\pi G}r_{0}^4\quad, \quad P=\frac{\Omega_{5}}{16\pi G}r_{0}^4~~, 
\end{equation}
agreeing with standard results \cite{Bhattacharyya:2007vjd, Banerjee_2011} when identifying $G_5=G/\Omega_5$ with $G_5$ being Newton's constant in five dimensions. In this limit, the spin current and chemical potential become
\begin{equation}\label{eq:D3spincurrent}
\begin{split}
     {\mathcal{S}_{a}}^{i j}=\frac{\Omega_{5}}{8\pi G}\ell u_{a}r_{0}^2\mathcal{H}^2 l^{A}\epsilon^{i j}_{A} ~~,~~\mu^{ij}=\frac{1}{2}\frac{r_{0}^2}{\mathcal{H}^2}\sum_{A=1}^{3}\frac{l_{A}}{(l_{A}^2+r_{H}^2)}\epsilon_{A}^{i j}~~,
\end{split}     
\end{equation}
where $\mathcal{H}$ is identified as the AdS radius $L$. We can see that the spin current is finite in the near-horizon limit and we can identify the spin density $s^{ij}=\Omega_{5}r_{0}^2\mathcal{H}^2 l^{A}\epsilon^{i j}_{A}/(8\pi G)$. The temperature and entropy take the form 
\begin{equation}
    T=\frac{4-2\sum_{A=1}^{3} \frac{l_{A}^2}{l_{A}^2+r_{H}^2}}{4\pi r_{H}}\frac{r_{0}^2}{\mathcal{H}^2}\quad,\quad s=\frac{\Omega_{5}}{4G}r_{0}^2\mathcal{H}^2r_{H}~,
\end{equation}
in the near-horizon limit. To match with the hydrodynamic theory in the main text we write the pressure $P$ in the grand canonical ensemble, that is, as a function of $T$ and $\mu^{ij}$. Explicitly, when all angular momenta are equal we can write $P=T^4f(\hat\mu)$ where
\begin{equation}
    f_{3}(\hat \mu)=\frac{\Omega_{5}L^{8}}{2^8\pi G}\hat{a}_{3}(\hat{\mu})^2\left(\hat{a}_{3}(\hat{\mu})^2+\frac{8}{3}\hat{\mu}^2\right)~,
\end{equation}
and where we defined $\hat \mu=\mu/T$ with $\mu=\sqrt{\mu_{ij}\mu^{ij}}$ and $\hat{a}=\pi+\sqrt{\pi^2+4\hat{\mu}^2/3}$. When the D3 brane is not spinning, then $\hat \mu=0$ implying $P=(1/8)N^2 \pi^2 T^4$ which is the correct result for $\mathcal{N}=4$ SYM at strong coupling. 

\vspace{0.5em}
\noindent
\textit{Derivative corrections in the near-horizon limit---}
Hydrodynamic corrections to the currents of equal spinning D3 branes were found in \cite{Erdmenger:2008rm, Banerjee:2008th, Son:2009tf} and explicitly written down fully in \cite{Megias:2013joa} in 5d. Here we uplift these results to 10d following the prescription described in the main text. Throughout this section we set $L=1$. In particular, we trivially find that the correction to the stress tensor has a shear viscosity component
\begin{equation}
\mathcal{T}^{ab}=-\eta \sigma^{ab}+\mathcal{O}(\partial^2)~~,    
\end{equation}
where $\eta=\Omega_ 5 r_0^2 r_{H}/(8\pi G)$ and no bulk viscosity due to conformal symmetry. From \cite{Megias:2013joa} the (3 copies of the) R-current read
\begin{equation}
\begin{split}
    J_{a}^{A}=n^{A} u_{a}-{\sigma^{A}}_B \left(T {P_a}^{b}\partial_b\left(\frac{\mu^{B}}{T}\right)-E^{B}_{a}\right)+\xi^{A}_M \omega_{a}+\xi^{AB}_{M} B_{B a}+\mathcal{O}(\partial^2)~,
\end{split}
\end{equation}\\
where the electric and magnetic fields follow from $E^{A}_{a}=u^{b}F_{ab}^A$ and $B^{a A}=(1/2)\epsilon^{a b c d}u_{b}F^{A}_{c d}$. We may uplift this expression to 10d invoking the dualities $\ell J_{a}^{A}\to {\mathcal{S}_{a}}^{ij}$ and $F_{ab}^B\to 2\ell {\Omega_{ab}}^{ij}$. To evaluate the various coefficients in \eqref{eq:correctedcurrents} we also use the translation of thermodynamic quantities from 5d to 10d, for instance the relation between the 5d horizon $r_{+}$ and the 10d horizon $r_{H}$, namely $r_{+}^3=r_{0}^2r_{H}$, as well as the translation of transport coefficients, for instance   $\mathcal{D}_{ijkl}\to T \sigma_{AB}$. The transport coefficients appearing in the second equation in \eqref{eq:correctedcurrents} then read
\begin{equation}
    \mathcal{D}_{ijkl}=\frac{\pi \Omega_{5} r_{H}^{\frac{7}{3}}T^3}{48 G r_{0}^{\frac{10}{3}}}\sum_{A,B=1}^{3}\epsilon_{ij}^{A}\epsilon_{kl}^{B} \quad,\quad \xi_{ij}=2l^2C\sum_{A=1}^{3}\epsilon_{ij}^{A}\quad,\quad \xi_{ijkl}^{S}=lC\frac{r_{0}^{\frac{2}{3}}(1+3r_{0}^{-\frac{4}{3}}r_{H}^{\frac{4}{3}})}{2r_{H}^{\frac{2}{3}}}\sum_{A,B=1}^{3}\epsilon_{ij}^{A}\epsilon_{kl}^{B}~~,
\end{equation}
where $l=l_{1}=l_{2}=l_{3}$ is the only angular momentum parameter in the equally spinning case and $C=C_{123456}$ is the only independent component of the 10d anomaly coefficient.


\end{document}